\documentclass[
 aps,
 preprint,
 fleqn
]{revtex4}

\usepackage[T1]{fontenc}
\usepackage{amsmath}
\usepackage{amssymb}
\usepackage{amsfonts}

\newcommand{\ME}[3]{\langle {#1} | {#2} | {#3} \rangle}
\newcommand{\KET}[1]{| {#1} \rangle}

\newcommand{\lslash}{\rlap{$\displaystyle \hspace{-0.7ex} \diagup $}l}
\newcommand{\llslash}{\rlap{$\displaystyle \hspace{-0.7ex} \diagup $}l'}
\newcommand{\sslash}{\rlap{$\displaystyle \hspace{-0.4ex} \diagup $}s}
\newcommand{\qslash}{\rlap{$\displaystyle \hspace{-0.4ex} \diagup $}q}
\newcommand{\kslash}{\rlap{$\displaystyle \hspace{-0.5ex} \diagup $}k}

\newcommand{\PPslash}{\rlap{$\displaystyle \hspace{-0.4ex} \diagup $}P'}

\DeclareMathOperator{\intd}{d}

\DeclareMathOperator{\Tr}{Tr}
\DeclareMathOperator{\sign}{sign}

\begin{document}

\title{Contribution of Hard Near Threshold Pion Electro Production to Nucleon Structure Functions}
\date{\today}
\author{Thorsten Sachs}
\affiliation{Institut für Theoretische Physik II, Ruhr-Universität Bochum, D-44780 Bochum, Germany}

\begin{abstract}
We provide a precise calculation for the contribution of hard near threshold pion electro production to all nucleon structure functions. This framework will allow us to split the required nucleon to pion nucleon transition probability matrix element into an exact threshold part parametrized by nucleon to pion nucleon transition form factors plus soft pion structures. These transition form factors will be related to nucleon form factors by taking into account symmetric and antisymmetric contributions of the needed leading twist nucleon distribution amplitude. Whereas the relations generated by the symmetric contributions are already known, this work will improve these relations by adding reasonable antisymmetric contributions. The obtained results show enhanced and significant consistency to experimental data, so that various predictions for further interesting experiments will be presented.
\end{abstract}

\maketitle

\section{Introduction}

Near threshold pion electro production has been studied for many years. Using soft pion theorems, the desired transition probability matrix elements were evaluated at low virtualities of the exchanged photon. One can find low energy theorems in the approximation of the vanishing pion mass in \cite{Kroll:1954}, \cite{Nambu:1962a}, \cite{Nambu:1962b} and for finite pion mass corrections in \cite{Vainshtein:1972} and \cite{Scherer:1991}. The expansion at small photon virtualities together with negligible pion mass has to be done with care because these limits do not commute, in general. All these theorems do not work at large virtualities, as discussed in \cite{Pobylitsa:2001}. In case of asymptotically large virtuality, one can use the QCD factorization theorem, see \cite{Lepage:1979a}, \cite{Lepage:1979b}, \cite{Lepage:1980}, and \cite{Efremov:1980a}, \cite{Efremov:1980b}, \cite{Efremov:1981}. So the representations obtained in \cite{Pobylitsa:2001} can be derived. Applying this technique, we start from asymptotically large virtuality and go to reasonable values systematically. Another approach is studied in \cite{Braun:2007} for extensive applications. Hereby, the area of $5-10\,\textnormal{GeV}^2$ is considered. Starting from the low energy behavior, the desired process is also discussed in \cite{Braun:2008}. In this work, the region of $1-10\,\textnormal{GeV}^2$ is considered.

First of all, we will evaluate the cross section. Afterwards, we will determine the nucleon structure functions. At next, we will discuss the expansion of the transition form factors. Finally, we will compare our results with experimental data.

\section{Evaluation of the Cross Section}

Let us start with the general representation of the scattering process
\pagebreak[0]
\begin{equation}
e(l,s)+N_i(P,S) \longrightarrow e(l',s')+N_f(P',S')+\pi^a(k).
\end{equation}

The momenta and spins of the particles are introduced in this expression and corresponding masses are denoted by $m_e$, $m_N$ and $m_\pi$. Possible combinations of nucleons and pions are constrained by charge conservation.

These processes will be considered at large virtuality $Q^2=-q^2=-(l-l')^2$ of the exchanged photon and near the production threshold of the pion, meaning that $W^2=(P'+k)^2=(P+q)^2$ is close to $W_{th}^2=(m_N+m_{\pi})^2$. We will often use the center of mass frame concerning the nucleon to pion nucleon transition subprocess, given by $\vec{P'}+\vec{k}=\vec{P}+\vec{q}=0$. This frame will allow to derive formulas for all required kinematic variables which will depend on $Q^2$ and $W^2$ only.

The discussed scenario is described by a scattering process of two particles in the initial states and three particles in the final states. Therefore, we have to study the following general formula for the cross section
\pagebreak[0]
\begin{equation}
\intd\!\sigma = \frac{|\mathcal{M}|^2}{4\sqrt{(P\cdot l)^2-m_N^2 m_e^2}}(2\pi)^4\delta^{(4)}(l'+P'+k-l-P)
\frac{\intd^3\!\vec{l'}}{(2\pi)^3 2l'_0}\frac{\intd^3\!\vec{P'}}{(2\pi)^3 2P'_0}\frac{\intd^3\!\vec{k}}{(2\pi)^3 2k_0}.
\end{equation}

Our aim is to derive an expression for this cross section which is compatible to the cross section representation of inclusive inelastic electron nucleon scattering.

The integration over the hadronic parts require the specified center of mass frame. We end up with the scattering angle of the pion $\Omega_\pi$ and the component $|\vec{k}|$ expressible as function of $W^2$ only. By convention, we introduce a function called $\beta(W)$ in order to describe this component under the condition $2|\vec{k}|=W\beta(W)$, noting that $\beta(W_{th})=0$,
\pagebreak[0]
\begin{equation}
\beta(W)=\sqrt{1-\frac{(m_N+m_\pi)^2}{W^2}}\sqrt{1-\frac{(m_N-m_\pi)^2}{W^2}}.
\end{equation}

The integration over the leptonic part requires the laboratory frame, $\vec{P}=0$. We label the energy of the initial lepton with $E_{lab}$ and the energy of the final lepton with $E'_{lab}$ in this frame. The scattering angle of the final lepton is denoted by $\Omega'_{lab}$. Neglecting the electron mass in the flux factor, one gets an intermediate result for the cross section
\pagebreak[0]
\begin{equation}
\label{c1}
\frac{\intd\!\sigma}{\intd\!E'_{lab}\intd\!\Omega'_{lab}}=
\int \left( \frac{E'_{lab}}{E_{lab}} \right) \frac{\beta(W)\intd\!\Omega_\pi}{2(4\pi)^5m_N}|\mathcal{M}|^2.
\end{equation}

The scattering amplitude contribution can be written as $|\mathcal{M}|^2=((4\pi\alpha_{em})^2/Q^4)L^{\mu\nu}M_{\mu\nu}$. Hereby we introduced the leptonic tensor $L^{\mu\nu}=(\bar{u}(l',s')\gamma^\mu u(l,s))(\bar{u}(l',s')\gamma^\nu u(l,s))^\dag$. Concerning this special process, the hadronic transition tensor is given by the structure $M_{\mu\nu}=(\ME{N_f(P',S')\pi^a(k)}{J_\mu^{em}(0)}{N_i(P,S)})(\ME{N_f(P',S')\pi^a(k)}{J_\nu^{em}(0)}{N_i(P,S)})^\dag$. This tensor is related to another hadronic transition tensor as follows
\pagebreak[0]
\begin{equation}
W_{\mu\nu}=\int\frac{\beta(W)\intd\!\Omega_\pi}{2(4\pi)^3m_N}M_{\mu\nu}.
\end{equation}

Inserting these structures into (\ref{c1}), one gets the final result for the cross section
\pagebreak[0]
\begin{equation}
\label{c2}
\frac{\intd\!\sigma}{\intd\!E'_{lab}\intd\!\Omega'_{lab}}=
\int \left( \frac{E'_{lab}}{E_{lab}} \right) \frac{\alpha_{em}^2}{Q^4}L^{\mu\nu}W_{\mu\nu}.
\end{equation}

It is noteworthy that we have to fix the initial state and to sum over all possible final states for the nucleon to pion nucleon transition. Concerning the spins of the initial particles, we have to distinguish between the unpolarized and the polarized case. Additionally, we have to sum over the spins of all final
particles.

The derived result for the cross section given in (\ref{c2}) is identical to the cross section representation of inclusive inelastic electron nucleon scattering and we already know from this process that the hadronic transition tensor $W_{\mu\nu}$ can be parameterized in nucleon structure functions. This important connection will allow us to extract results for the nucleon structure functions finally.

In the case of an unpolarized nucleon target, $W_{\mu\nu}$ can be parameterized by two nucleon structure functions $F_1$ and $F_2$. This parametrization is symmetric under the interchange of $\mu\leftrightarrow\nu$ and according to this, we call this tensor $W_{\mu\nu}^S$. One gets
\pagebreak[0]
\begin{equation}
\label{s}
W_{\mu\nu}^S=\frac{1}{m_N}\bigg[F_1(W,Q^2)(\frac{q_\mu q_\nu}{q^2}-g_{\mu\nu})+
\frac{1}{P\cdot q}F_2(W,Q^2)(P_\mu-\frac{P\cdot q}{q^2}q_\mu)(P_\nu-\frac{P\cdot q}{q^2}q_\nu)\bigg].
\end{equation}

In the case of a polarized nucleon target, we have to deal with an antisymmetric parametrization $W_{\mu\nu}^A$ additional to the already introduced symmetric parametrization. This contribution can be expressed by two nucleon structure functions $G_1$ and $G_2$. We obtain
\pagebreak[0]
\begin{equation}
\label{a}
W_{\mu\nu}^A=i\varepsilon_{\mu\nu\rho\sigma}\bigg[\frac{1}{P\cdot q}G_1(W,Q^2)q^\rho S^\sigma+
\frac{1}{P\cdot q}G_2(W,Q^2)(q^\rho S^\sigma-\frac{S\cdot q}{P\cdot q}q^\rho P^\sigma)\bigg].
\end{equation}

Finally, we have to consider the required nucleon to pion nucleon transition probability matrix element given by $\ME{N_f(P',S')\pi^a(k)}{J_\mu^{em}(0)}{N_i(P,S)}$ in order to calculate the hadronic transition tensor. We need a representation of this matrix element at large momentum transfer and near the production threshold of the pion.

At the exact threshold limit, one can express it by an expansion in form factors similar to usual nucleon transition probability matrix elements. We designate the nucleon to pion nucleon transition form factors by $\mathcal{A}(\gamma^\star N_i \rightarrow N_f \pi^a)$, see \cite{Pobylitsa:2001}. The discussed expression must respect the current conservation relation $q^\mu\ME{N_f(P',S')\pi^a(k)}{J_\mu^{em}(0)}{N_i(P,S)}=0$, realizable by the replacement
$\gamma_\mu \rightarrow \gamma_\mu-(\qslash/q^2)q_\mu$. One obtains the S-wave part of the process
\pagebreak[0]
\begin{equation}
\textnormal{S-wave}=\mathcal{A}(\gamma^\star N_i \rightarrow N_f \pi^a)(Q^2)\bar{N}(P',S')(\gamma_\mu-\frac{\qslash}{q^2}q_\mu)\gamma_5N(P,S).
\end{equation}

In the near threshold region, we can use soft pion theorems. Additional to the already considered contribution for the exact threshold situation, we get pole contributions known from soft pion theorems. The emission of the pion from the final nucleon generates a large contribution for $W^2$ close to $W_{th}^2$ and therefore the dominant contribution. The emission of the pion from the initial nucleon is suppressed at large $Q^2$ and one can neglect it consequentially. Additional to these both nucleon poles, we obtain a pion pole which can also be neglected at large $Q^2$. The expression for the dominant contribution was calculated under the constraint that it vanishes at the exact threshold. Moreover, one has to apply the specified current conservation replacement again, so we get the P-wave part of the process
\pagebreak[0]
\begin{equation}
\textnormal{P-wave}=\frac{ig_A\tau_{fi}^aG_M^{N_i}(Q^2)}{2f_\pi(W^2-m_N^2)}\bar{N}(P',S')\kslash\gamma_5(\PPslash+m_N)(\gamma_\mu-\frac{\qslash}{q^2}q_\mu)N(P,S).
\end{equation}

\section{Determination of the Nucleon Structure Functions}

Our aim is to calculate the leptonic tensor and the hadronic transition tensor. The result of the leptonic tensor is already known, but we will present its derivation in order to introduce the required technique. The calculation of both tensors require the summation over the spins of the final particles. Concerning the spins of the initial particles, we have to distinguish between the unpolarized and the polarized case. The unpolarized scenario requires the average over the spins of the initial particles and for the polarized case, we have to express the spins of the initial particles by a corresponding covariant spin vector. We realized that the hadronic transition tensor can be parameterized in terms of nucleon structure functions and we will determine them finally. For their computation, we have to fix the initial state and to sum over all possible final states. Therefore, we define $X^p=\{ p\pi^0,n\pi^+ \}$ and
$X^n=\{ n\pi^0,p\pi^- \}$ as the possible transition states for the proton and the neutron. Finally, we will consider the discussed limit of large momentum transfer and small invariant masses, meaning that $x_B=Q^2/(2(P\cdot q))=Q^2/(Q^2+W^2-m_N^2)$ is close to the threshold value $x_B=[(1+(2m_N+m_\pi)m_\pi)/Q^2]^{-1}$. For large $Q^2$ one gets $x_B \rightarrow 1$ consequentially.

\subsection{Leptonic Tensor}

At first, we calculate the leptonic tensor for unpolarized electrons
\pagebreak[0]
\begin{equation}
L^{\mu\nu}=\frac{1}{2}\sum_{s=\uparrow,\downarrow}\sum_{s'=\uparrow,\downarrow}(\bar{u}(l',s')\gamma^\mu u(l,s))(\bar{u}(l',s')\gamma^\nu u(l,s))^\dag =\frac{1}{2}\Tr[(\llslash+m_e)\gamma^\mu(\lslash+m_e)\gamma^\nu].
\end{equation}
One can see that this trace structure is symmetric under the interchange of $\mu\leftrightarrow\nu$ and according to this, we can write $L^{\mu\nu}=L^{\mu\nu}_S$. The result is given by
\pagebreak[0]
\begin{equation}
L^{\mu\nu}_S=2(l^\mu l'^\nu+l'^\mu l^\nu)+q^2g^{\mu\nu}.
\end{equation}

At next, we calculate the leptonic tensor for polarized electrons
\pagebreak[0]
\begin{equation}
L^{\mu\nu}=\sum_{s'=\uparrow,\downarrow}(\bar{u}(l',s')\gamma^\mu u(l,s))(\bar{u}(l',s')\gamma^\nu u(l,s))^\dag
=\Tr[(\llslash+m_e)\gamma^\mu(\lslash+m_e)\frac{1+\gamma_5\sslash}{2}\gamma^\nu].
\end{equation}
This trace structure is the sum of a symmetric and an antisymmetric part and the symmetric part is identical to $L^{\mu\nu}_S$. We can use this property and write $L^{\mu\nu}=L^{\mu\nu}_S+L^{\mu\nu}_A$. Consequentially, we just need to compute the antisymmetric trace structure
\pagebreak[0]
\begin{equation}
L^{\mu\nu}_A=-2im_e\varepsilon^{\mu\nu\rho\sigma}q_\rho s_\sigma.
\end{equation}

One can expect the same symmetry properties for the hadronic transition tensor as obtained for the leptonic tensor.

\subsection{Unpolarized Hadronic Transition Tensor}

The unpolarized hadronic transition tensor can be written as sum of three components corresponding to the S-wave contribution, the P-wave contribution and the interference term. When we transform these components in trace structures, one can see that both unmixed components are symmetric whereas the mixed component is antisymmetric.

The calculation of the symmetric S-wave contribution is straightforward and delivers the following result
\pagebreak[0]
\begin{equation}
\begin{split}
& W_{\mu\nu}^{S(S)}=\int\frac{\beta(W)\intd\!\Omega_\pi}{(4\pi)^3m_N}|\mathcal{A}(\gamma^\star N_i \rightarrow N_f \pi^a)|^2 \\
& \bigg[((P\cdot P')+m_N^2)(\frac{q_\mu q_\nu}{q^2}-g_{\mu\nu})+P_\mu P'_\nu+P'_\mu P_\nu \\
& -\frac{P'\cdot q}{q^2}(P_\mu q_\nu+q_\mu P_\nu)-\frac{P\cdot q}{q^2}(P'_\mu q_\nu+q_\mu P'_\nu)+\frac{2(P\cdot q)(P'\cdot q)}{q^4}q_\mu q_\nu\bigg].
\end{split}
\end{equation}

The calculation of the symmetric P-wave contribution requires the center of mass frame relation $4((P'\cdot k)^2-m_N^2k^2)=W^4\beta^2(W)$. One gets the following expression
\pagebreak[0]
\begin{equation}
\begin{split}
& W_{\mu\nu}^{S(P)}=\int\frac{W^4\beta^3(W)\intd\!\Omega_\pi}{(4\pi)^3m_N}\frac{g_A^2(\tau_{fi}^a)^2(G_M^{N_i}(Q^2))^2}{4f_\pi^2(W^2-m_N^2)^2} \\
& \bigg[((P\cdot P')-m_N^2)(\frac{q_\mu q_\nu}{q^2}-g_{\mu\nu})+P_\mu P'_\nu+P'_\mu P_\nu \\
& -\frac{P'\cdot q}{q^2}(P_\mu q_\nu+q_\mu P_\nu)-\frac{P\cdot q}{q^2}(P'_\mu q_\nu+q_\mu P'_\nu)+\frac{2(P\cdot q)(P'\cdot q)}{q^4}q_\mu q_\nu\bigg].
\end{split}
\end{equation}

Furthermore, we have to consider the antisymmetric interference term. This part can be represented as follows
\pagebreak[0]
\begin{equation}
\begin{split}
& W_{\mu\nu}^{A(I)}=\int\frac{\beta(W)\intd\!\Omega_\pi}{(4\pi)^3}\frac{2im_Ng_A\tau_{fi}^aG_M^{N_i}(Q^2)}{f_\pi(W^2-m_N^2)}
\mathcal{A}(\gamma^\star N_i \rightarrow N_f \pi^a) \\
& \bigg[\frac{P'\cdot q}{q^2}(q_\mu k_\nu-k_\mu q_\nu)+\frac{q\cdot k}{q^2}(P'_\mu q_\nu-q_\mu P'_\nu)-P'_\mu k_\nu+k_\mu P'_\nu\bigg].
\end{split}
\end{equation}
This term should not contribute to the cross section, because we consider an unpolarized nucleon target. To prove this, we contract with the leptonic tensor and evaluate the remaining integration. The contraction with the symmetric leptonic tensor is obviously zero, but the contraction with the antisymmetric leptonic tensor delivers a non-vanishing result. We change from the four dimensional representation to the three dimensional representation by applying the center of mass frame
\pagebreak[0]
\begin{equation}
L^{\mu\nu}_A W_{\mu\nu}^{A(I)}=\int\frac{W\beta(W)\intd\!\Omega_\pi}{(4\pi)^3}\frac{8m_Nm_eg_A\tau_{fi}^aG_M^{N_i}(Q^2)}{f_\pi(W^2-m_N^2)}
\mathcal{A}(\gamma^\star N_i \rightarrow N_f \pi^a)\vec{k}\cdot(\vec{q}\times\vec{s}).
\end{equation}
The remaining integration delivers $\int\intd\!\Omega_\pi\vec{k}\cdot(\vec{q}\times\vec{s})=0$, leading to $L^{\mu\nu}_A W_{\mu\nu}^{A(I)}=0$.

The contributing parts of $W_{\mu\nu}$ are all symmetric, namely $W_{\mu\nu}^{S(S)}$ and $W_{\mu\nu}^{S(P)}$. In order to obtain the required tensor structures, we have to eliminate the dependence on $P'$ in a special way. Therefore, we have to apply the center of mass frame. In this case, one can evaluate the remaining integration and obtain for $P'_0$ the correlation $2W^2P'_0=(W^2+m_N^2-m_\pi^2)(P+q)$. Through this procedure, we can derive the required tensor structures for both components
\pagebreak[0]
\begin{equation}
\begin{split}
& W_{\mu\nu}^{S(S)}=\frac{\beta(W)}{(4\pi)^2m_N}|\mathcal{A}(\gamma^\star N_i \rightarrow N_f \pi^a)|^2 \\
& \bigg[(\frac{W^2+m_N^2-m_\pi^2}{2W^2}((P\cdot q)+m_N^2)+m_N^2)(\frac{q_\mu q_\nu}{q^2}-g_{\mu\nu}) \\
& +\frac{W^2+m_N^2-m_\pi^2}{W^2}(P_\mu-\frac{P\cdot q}{q^2}q_\mu)(P_\nu-\frac{P\cdot q}{q^2}q_\nu)\bigg]
\end{split}
\end{equation}
\begin{equation}
\begin{split}
& W_{\mu\nu}^{S(P)}=\frac{W^4\beta^3(W)}{(4\pi)^2m_N}\frac{g_A^2(\tau_{fi}^a)^2(G_M^{N_i}(Q^2))^2}{4f_\pi^2(W^2-m_N^2)^2} \\
& \bigg[(\frac{W^2+m_N^2-m_\pi^2}{2W^2}((P\cdot q)+m_N^2)-m_N^2)(\frac{q_\mu q_\nu}{q^2}-g_{\mu\nu}) \\
& +\frac{W^2+m_N^2-m_\pi^2}{W^2}(P_\mu-\frac{P\cdot q}{q^2}q_\mu)(P_\nu-\frac{P\cdot q}{q^2}q_\nu)\bigg].
\end{split}
\end{equation}

The expressions for the unpolarized nucleon structure functions can be extracted by comparison with (\ref{s}). Applying now the kinematic relation 
$2(P\cdot q)=(Q^2+W^2-m_N^2)$, one gets the desired dependence on $Q^2$ and $W^2$. We obtain
\pagebreak[0]
\begin{equation}
\begin{split}
& F_1^p(W,Q^2)= \\
& +((W^2+m_N^2-m_\pi^2)(Q^2+W^2+m_N^2)+4W^2m_N^2)\frac{\beta(W)}{4(4\pi)^2W^2}
\sum_{X=X^p}|\mathcal{A}(\gamma^\star p \rightarrow X)|^2 \\
& +((W^2+m_N^2-m_\pi^2)(Q^2+W^2+m_N^2)-4W^2m_N^2)
\frac{3g_A^2(G_M^p(Q^2))^2W^2\beta^3(W)}{16(4\pi)^2 f_\pi^2(W^2-m_N^2)^2}
\end{split}
\end{equation}
\begin{equation}
\begin{split}
& F_1^n(W,Q^2)= \\
& +((W^2+m_N^2-m_\pi^2)(Q^2+W^2+m_N^2)+4W^2m_N^2)\frac{\beta(W)}{4(4\pi)^2W^2}
\sum_{X=X^n}|\mathcal{A}(\gamma^\star n \rightarrow X)|^2 \\
& +((W^2+m_N^2-m_\pi^2)(Q^2+W^2+m_N^2)-4W^2m_N^2)
\frac{3g_A^2(G_M^n(Q^2))^2W^2\beta^3(W)}{16(4\pi)^2 f_\pi^2(W^2-m_N^2)^2}
\end{split}
\end{equation}
\begin{equation}
\begin{split}
& F_2^p(W,Q^2)= \\
& +(W^2+m_N^2-m_\pi^2)(Q^2+W^2-m_N^2)\frac{\beta(W)}{2(4\pi)^2W^2}
\sum_{X=X^p}|\mathcal{A}(\gamma^\star p \rightarrow X)|^2 \\
& +(W^2+m_N^2-m_\pi^2)(Q^2+W^2-m_N^2)
\frac{3g_A^2(G_M^p(Q^2))^2W^2\beta^3(W)}{8(4\pi)^2f_\pi^2(W^2-m_N^2)^2}
\end{split}
\end{equation}
\begin{equation}
\begin{split}
& F_2^n(W,Q^2)= \\
& +(W^2+m_N^2-m_\pi^2)(Q^2+W^2-m_N^2)\frac{\beta(W)}{2(4\pi)^2W^2}
\sum_{X=X^n}|\mathcal{A}(\gamma^\star n \rightarrow X)|^2 \\
& +(W^2+m_N^2-m_\pi^2)(Q^2+W^2-m_N^2)
\frac{3g_A^2(G_M^n(Q^2))^2W^2\beta^3(W)}{8(4\pi)^2f_\pi^2(W^2-m_N^2)^2}.
\end{split}
\end{equation}
Finally, we consider the case when $Q^2$ is large in comparison with all appearing hadron masses and $W^2$ is close to $W_{th}^2$. At this limit, one gets the following results for the unpolarized nucleon structure functions
\pagebreak[0]
\begin{eqnarray}
F_1^p(W,Q^2) &=& \frac{Q^2\beta(W)}{2(4\pi)^2}\left[\sum_{X=X^p}|\mathcal{A}(\gamma^\star p \rightarrow X)|^2+
\frac{3g_A^2(G_M^p(Q^2))^2W^4\beta^2(W)}{4f_\pi^2(W^2-m_N^2)^2}\right] \\
F_1^n(W,Q^2) &=& \frac{Q^2\beta(W)}{2(4\pi)^2}\left[\sum_{X=X^n}|\mathcal{A}(\gamma^\star n \rightarrow X)|^2+
\frac{3g_A^2(G_M^n(Q^2))^2W^4\beta^2(W)}{4f_\pi^2(W^2-m_N^2)^2}\right]
\end{eqnarray}
\begin{eqnarray}
F_2^p(W,Q^2) &=& \frac{Q^2\beta(W)}{(4\pi)^2}\left[\sum_{X=X^p}|\mathcal{A}(\gamma^\star p \rightarrow X)|^2+
\frac{3g_A^2(G_M^p(Q^2))^2W^4\beta^2(W)}{4f_\pi^2(W^2-m_N^2)^2}\right] \\
F_2^n(W,Q^2) &=& \frac{Q^2\beta(W)}{(4\pi)^2}\left[\sum_{X=X^n}|\mathcal{A}(\gamma^\star n \rightarrow X)|^2+
\frac{3g_A^2(G_M^n(Q^2))^2W^4\beta^2(W)}{4f_\pi^2(W^2-m_N^2)^2}\right].
\end{eqnarray}
The important remark here is that the result for the unpolarized nucleon structure function $F_2(W,Q^2)$ for a specified nucleon is precisely a factor of two larger than the result for the unpolarized nucleon structure function $F_1(W,Q^2)$ for the same nucleon, which is in agreement with the parton model prediction $F_2=2x_BF_1$ applying the limit $x_B \rightarrow 1$.

\subsection{Polarized Hadronic Transition Tensor}

Similar to the unpolarized case, we can split the tensor into its wave components and transform these parts in trace structures. For both unmixed parts, we obtain a representation as sum of the known symmetric part and a new antisymmetric part. The mixed part can be written as sum of the known antisymmetric part and a new symmetric part. Analog to the calculation of the polarized leptonic tensor, we have to focus on the new parts only.

Concerning the antisymmetric S-wave component, one gets the following contribution to the polarized hadronic transition tensor directly
\pagebreak[0]
\begin{equation}
\begin{split}
& W_{\mu\nu}^{A(S)}=
i\varepsilon_{\mu\nu\rho\sigma}\int\frac{\beta(W)\intd\!\Omega_\pi}{(4\pi)^3}|\mathcal{A}(\gamma^\star N_i \rightarrow N_f \pi^a)|^2 \\
& \bigg[\frac{P'\cdot q}{q^2}q^\rho S^\sigma+\frac{P\cdot q}{q^2}q^\rho S^\sigma
-\frac{S\cdot q}{q^2}q^\rho P'^\sigma-\frac{S\cdot q}{q^2}q^\rho P^\sigma\bigg].
\end{split}
\end{equation}

Concerning the antisymmetric P-wave component, one needs the center of mass relation introduced in the corresponding case before. We obtain the following contribution to the polarized hadronic transition tensor
\pagebreak[0]
\begin{equation}
\begin{split}
& W_{\mu\nu}^{A(P)}=i\varepsilon_{\mu\nu\rho\sigma}\int\frac{W^4\beta^3(W)\intd\!\Omega_\pi}{(4\pi)^3}
\frac{g_A^2(\tau_{fi}^a)^2(G_M^{N_i}(Q^2))^2}{4f_\pi^2(W^2-m_N^2)^2} \\
& \bigg[\frac{P'\cdot q}{q^2}q^\rho S^\sigma-\frac{P\cdot q}{q^2}q^\rho S^\sigma
-\frac{S\cdot q}{q^2}q^\rho P'^\sigma+\frac{S\cdot q}{q^2}q^\rho P^\sigma\bigg].
\end{split}
\end{equation}

Moreover, we have to calculate the symmetric interference term. Applying the Schouten identity frequently, one receives the following result
\pagebreak[0]
\begin{equation}
\begin{split}
& W_{\mu\nu}^{S(I)}=\int\frac{\beta(W)\intd\!\Omega_\pi}{(4\pi)^3}\frac{2g_A\tau_{fi}^aG_M^{N_i}(Q^2)}{f_\pi(W^2-m_N^2)}
\mathcal{A}(\gamma^\star N_i \rightarrow N_f \pi^a) \\
& \bigg[\bigg(\frac{2(P\cdot q)}{q^4}q_\mu q_\nu-\frac{1}{q^2}(P_\mu q_\nu+q_\mu P_\nu-q_\mu q_\nu)-g_{\mu\nu}\bigg)
\varepsilon_{\alpha\beta\gamma\delta}P'^\alpha k^\beta P^\gamma S^\delta \\
& -\bigg(\frac{P\cdot q}{q^2}q_\mu-P_\mu\bigg)\varepsilon_{\nu\beta\gamma\delta}k^\beta P^\gamma S^\delta
-\bigg(\frac{P\cdot q}{q^2}q_\nu-P_\nu\bigg)\varepsilon_{\mu\beta\gamma\delta}k^\beta P^\gamma S^\delta \\
& +\bigg(\frac{q\cdot k}{q^2}q_\mu-k_\mu\bigg)\varepsilon_{\nu\beta\gamma\delta}q^\beta P^\gamma S^\delta
+\bigg(\frac{q\cdot k}{q^2}q_\nu-k_\nu\bigg)\varepsilon_{\mu\beta\gamma\delta}q^\beta P^\gamma S^\delta\bigg].
\end{split}
\end{equation}
This part should not contribute to the cross section, because we consider a polarized nucleon target. The difference compared to the unpolarized case is that we get a vanishing result for the contraction with the antisymmetric leptonic tensor and a non-vanishing result for the contraction with the symmetric leptonic tensor
\pagebreak[0]
\begin{equation}
\begin{split}
& L^{\mu\nu}_S W_{\mu\nu}^{S(I)}=\int\frac{\beta(W)\intd\!\Omega_\pi}{(4\pi)^3}\frac{8g_A\tau_{fi}^aG_M^{N_i}(Q^2)}{f_\pi(W^2-m_N^2)}
\mathcal{A}(\gamma^\star N_i \rightarrow N_f \pi^a) \\
& \bigg[\bigg(\frac{1}{q^2}((P\cdot q)(q\cdot l')+(P\cdot l)(q\cdot l')+(P\cdot l')(q\cdot l)-(q\cdot l)(q\cdot l')) \\
& -\frac{2}{q^4}(P\cdot q)(q\cdot l)(q\cdot l')-P\cdot l'+l\cdot l'+\frac{1}{4}q^2\bigg)
(P'_0+k_0)(\vec{k}\cdot(\vec{P}\times\vec{S})) \\
& -l'_0(P\cdot(l+l'))(\vec{k}\cdot(\vec{P}\times\vec{S}))+P_0(P\cdot(l+l'))(\vec{k}\cdot(\vec{l'}\times\vec{S})) \\
& +(P'_0+k_0)(\vec{k}\cdot(\vec{l}+\vec{l'}))(\vec{l'}\cdot(\vec{P}\times\vec{S}))\bigg].
\end{split}
\end{equation}
The remaining integration delivers the same type of integrals as in the unpolarized case and we receive $L^{\mu\nu}_S W_{\mu\nu}^{S(I)}=0$ according to this.

Every contributing structure of $W_{\mu\nu}$ can be written as sum of an already known symmetric part and a new antisymmetric part and we only have to focus on the antisymmetric parts, namely $W_{\mu\nu}^{A(S)}$ and $W_{\mu\nu}^{A(P)}$. Applying the same technique as in the unpolarized case, we end up with the required tensor structures for both components
\pagebreak[0]
\begin{equation}
\begin{split}
& W_{\mu\nu}^{A(S)}=i\varepsilon_{\mu\nu\rho\sigma}\frac{\beta(W)}{(4\pi)^2}|\mathcal{A}(\gamma^\star N_i \rightarrow N_f \pi^a)|^2 \\
& \bigg[\frac{W^2+m_N^2-m_\pi^2}{2W^2}q^\rho S^\sigma+
\frac{3W^2+m_N^2-m_\pi^2}{2W^2}\frac{P\cdot q}{q^2}(q^\rho S^\sigma-\frac{S\cdot q}{P\cdot q}q^\rho P^\sigma)\bigg]
\end{split}
\end{equation}
\begin{equation}
\begin{split}
& W_{\mu\nu}^{A(P)}=i\varepsilon_{\mu\nu\rho\sigma}\frac{W^4\beta^3(W)}{(4\pi)^2}\frac{g_A^2(\tau_{fi}^a)^2(G_M^{N_i}(Q^2))^2}{4f_\pi^2(W^2-m_N^2)^2} \\
& \bigg[\frac{W^2+m_N^2-m_\pi^2}{2W^2}q^\rho S^\sigma+
\frac{-W^2+m_N^2-m_\pi^2}{2W^2}\frac{P\cdot q}{q^2}(q^\rho S^\sigma-\frac{S\cdot q}{P\cdot q}q^\rho P^\sigma)\bigg].
\end{split}
\end{equation}

The expressions for the polarized nucleon structure functions can be extracted by comparison with (\ref{a}). At this point, we have to apply the same kinematic relation as used in the unpolarized case before. According to this, we get
\pagebreak[0]
\begin{equation}
\begin{split}
& G_1^p(W,Q^2)= \\
& +(W^2+m_N^2-m_\pi^2)(Q^2+W^2-m_N^2)\frac{\beta(W)}{4(4\pi)^2W^2}
\sum_{X=X^p}|\mathcal{A}(\gamma^\star p \rightarrow X)|^2 \\
& +(W^2+m_N^2-m_\pi^2)(Q^2+W^2-m_N^2)
\frac{3g_A^2(G_M^p(Q^2))^2W^2\beta^3(W)}{16(4\pi)^2 f_\pi^2(W^2-m_N^2)^2}
\end{split}
\end{equation}
\begin{equation}
\begin{split}
& G_1^n(W,Q^2)= \\
& +(W^2+m_N^2-m_\pi^2)(Q^2+W^2-m_N^2)\frac{\beta(W)}{4(4\pi)^2W^2}
\sum_{X=X^n}|\mathcal{A}(\gamma^\star n \rightarrow X)|^2 \\
& +(W^2+m_N^2-m_\pi^2)(Q^2+W^2-m_N^2)
\frac{3g_A^2(G_M^n(Q^2))^2W^2\beta^3(W)}{16(4\pi)^2 f_\pi^2(W^2-m_N^2)^2}
\end{split}
\end{equation}
\begin{equation}
\begin{split}
& G_2^p(W,Q^2)= \\
& -(3W^2+m_N^2-m_\pi^2)(Q^2+W^2-m_N^2)^2\frac{\beta(W)}{8(4\pi)^2W^2Q^2}
\sum_{X=X^p}|\mathcal{A}(\gamma^\star p \rightarrow X)|^2 \\
& -(-W^2+m_N^2-m_\pi^2)(Q^2+W^2-m_N^2)^2
\frac{3g_A^2(G_M^p(Q^2))^2W^2\beta^3(W)}{32(4\pi)^2f_\pi^2(W^2-m_N^2)^2Q^2}
\end{split}
\end{equation}
\begin{equation}
\begin{split}
& G_2^n(W,Q^2)= \\
& -(3W^2+m_N^2-m_\pi^2)(Q^2+W^2-m_N^2)^2\frac{\beta(W)}{8(4\pi)^2W^2Q^2}
\sum_{X=X^n}|\mathcal{A}(\gamma^\star n \rightarrow X)|^2 \\
& -(-W^2+m_N^2-m_\pi^2)(Q^2+W^2-m_N^2)^2
\frac{3g_A^2(G_M^n(Q^2))^2W^2\beta^3(W)}{32(4\pi)^2f_\pi^2(W^2-m_N^2)^2Q^2}.
\end{split}
\end{equation}
We gain the following results for the polarized nucleon structure functions by considering the same limit as in the unpolarized case
\pagebreak[0]
\begin{eqnarray}
G_1^p(W,Q^2) &=& \frac{Q^2\beta(W)}{2(4\pi)^2}\left[\sum_{X=X^p}|\mathcal{A}(\gamma^\star p \rightarrow X)|^2+
\frac{3g_A^2(G_M^p(Q^2))^2W^4\beta^2(W)}{4f_\pi^2(W^2-m_N^2)^2}\right] \\
G_1^n(W,Q^2) &=& \frac{Q^2\beta(W)}{2(4\pi)^2}\left[\sum_{X=X^n}|\mathcal{A}(\gamma^\star n \rightarrow X)|^2+
\frac{3g_A^2(G_M^n(Q^2))^2W^4\beta^2(W)}{4f_\pi^2(W^2-m_N^2)^2}\right]
\end{eqnarray}
\begin{eqnarray}
G_2^p(W,Q^2) &=& -\frac{Q^2\beta(W)}{2(4\pi)^2}\left[\sum_{X=X^p}|\mathcal{A}(\gamma^\star p \rightarrow X)|^2\right] \\
G_2^n(W,Q^2) &=& -\frac{Q^2\beta(W)}{2(4\pi)^2}\left[\sum_{X=X^n}|\mathcal{A}(\gamma^\star n \rightarrow X)|^2\right].
\end{eqnarray}
The important conclusion here is that we receive no P-wave contribution for $G_2(W,Q^2)$. We mention that this contribution is not exactly zero, but it is suppressed at the considered limit. Despite from that, we notice the opposite sign in the final results.

\section{Transition Form Factor Expansion}

Our final aim is to derive expressions for the nucleon to pion nucleon transition form factors as functions of usual nucleon form factors at large $Q^2$. It should be clear that we need additional assumptions in order to obtain these connections. The main tool is the QCD factorization theorem allowing us to split the form factors in one hard part and two soft parts which can be calculated separately. It is recommended to have a look on \cite{Lepage:1979a}, \cite{Lepage:1979b}, \cite{Lepage:1980} and also \cite{Efremov:1980a}, \cite{Efremov:1980b}, \cite{Efremov:1981} for more details. By convention, we work with a spin configuration of $\uparrow\downarrow\uparrow$ for the quarks in a corresponding nucleon with spin $\uparrow$ and apply the notation $f(x)=f(x_i)$ concerning the dependence on the quark momentum fractions.

We will introduce the expressions of the hard parts now. Their representations depend on three functions $T_i(x,y)$. It can be shown that these three functions are identical for every required current. We do not need the explicit results for these functions, however we have to use that $T_1$ and $T_3$ get identical by exchanging $x_1\leftrightarrow x_3$ and $y_1\leftrightarrow y_3$ in one of them
\pagebreak[0]
\begin{eqnarray}
T_H^{em}(x,y,Q^2) &=& \frac{16}{9}\frac{(4\pi\bar{\alpha}_s)^2}{Q^4}\sum_{i=1}^{3}e_i[T_i(x,y)+T_i(y,x)] \\
T_{H5}^{(V-)}(x,y,Q^2) &=& \frac{16}{9}\frac{(4\pi\bar{\alpha}_s)^2}{Q^4}\sum_{i=1}^{3}\sign(h_i)[I_{i-}T_i(x,y)+I_{i-}T_i(y,x)] \\
T_{H5}^{(V+)}(x,y,Q^2) &=& \frac{16}{9}\frac{(4\pi\bar{\alpha}_s)^2}{Q^4}\sum_{i=1}^{3}\sign(h_i)[I_{i+}T_i(x,y)+I_{i+}T_i(y,x)] \\
T_{H5}^{(S)}(x,y,Q^2) &=& \frac{16}{9}\frac{(4\pi\bar{\alpha}_s)^2}{Q^4}\sum_{i=1}^{3}\sign(h_i)[T_i(x,y)+T_i(y,x)].
\end{eqnarray}

At first, we introduced the hard part expression generated by the electromagnetic current, see \cite{Lepage:1979b} and \cite{Lepage:1980}. Further information can be taken from \cite{Carlson:1987a}. At next, we considered the isovector axial-vector current contributions by applying both transition processes \cite{Carlson:1986}. One can look for further applications in \cite{Carlson:1987c}. At last, we considered the isoscalar axial-vector current contribution for completeness \cite{Carlson:1987b}.

The electromagnetic charge operator $e_i$ determines the charge of quark $i$ and so we get $e_i\KET{u}=e_u\KET{u}=\frac{2}{3}\KET{u}$ and $e_i\KET{d}=e_d\KET{d}=-\frac{1}{3}\KET{d}$. The operator $I_{i-}$ denotes the isospin lowering operator of quark $i$ leading to $I_{i-}\KET{u}=I_{u-}\KET{u}=\KET{d}$ and $I_{i-}\KET{d}=I_{d-}\KET{d}=\KET{0}$. The operator $I_{i+}$ denotes the isospin raising operator of quark $i$ leading to $I_{i+}\KET{d}=I_{d+}\KET{d}=\KET{u}$ and $I_{i+}\KET{u}=I_{u+}\KET{u}=\KET{0}$. The $\sign$ of the helicity of quark $i$ is generated by the sign operator $\sign(h_i)$ leading to $\sign(h_i)\KET{\uparrow}=\sign(h_\uparrow)\KET{\uparrow}=+\KET{\uparrow}$ and
$\sign(h_i)\KET{\downarrow}=\sign(h_\downarrow)\KET{\downarrow}=-\KET{\downarrow}$.

At next, we will summarize the representations of the soft parts. The required leading twist nucleon distribution amplitude will be split in distribution functions $\phi_S(x)$ and $\phi_A(x)$ which are symmetric and antisymmetric respectively under the exchange of $x_1 \leftrightarrow x_3$. These components are related to quarks with parallel helicities, see \cite{Pobylitsa:2001}. We obtain the following list of functions related to the corresponding hadronic states
\pagebreak[0]
\begin{eqnarray}
\phi_p(x)=
&+& \frac{1}{\sqrt{6}}\phi_S(x)\KET{2u_{\uparrow}d_{\downarrow}u_{\uparrow}-u_{\uparrow}u_{\downarrow}d_{\uparrow}-d_{\uparrow}u_{\downarrow}u_{\uparrow}}
\nonumber \\
&+& \frac{1}{\sqrt{2}}\phi_A(x)\KET{u_{\uparrow}u_{\downarrow}d_{\uparrow}-d_{\uparrow}u_{\downarrow}u_{\uparrow}}
\end{eqnarray}
\begin{eqnarray}
\phi_n(x)=
&-& \frac{1}{\sqrt{6}}\phi_S(x)\KET{2d_{\uparrow}u_{\downarrow}d_{\uparrow}-d_{\uparrow}d_{\downarrow}u_{\uparrow}-u_{\uparrow}d_{\downarrow}d_{\uparrow}}
\nonumber \\
&-& \frac{1}{\sqrt{2}}\phi_A(x)\KET{d_{\uparrow}d_{\downarrow}u_{\uparrow}-u_{\uparrow}d_{\downarrow}d_{\uparrow}}
\end{eqnarray}
\begin{eqnarray}
\phi_{p\pi^0}(x)=
&+& \frac{1}{2\sqrt{6}f_\pi}\phi_S(x)\KET{6u_{\uparrow}d_{\downarrow}u_{\uparrow}+u_{\uparrow}u_{\downarrow}d_{\uparrow}+d_{\uparrow}u_{\downarrow}u_{\uparrow}}
\nonumber \\
&-& \frac{1}{2\sqrt{2}f_\pi}\phi_A(x)\KET{u_{\uparrow}u_{\downarrow}d_{\uparrow}-d_{\uparrow}u_{\downarrow}u_{\uparrow}}
\end{eqnarray}
\begin{eqnarray}
\phi_{n\pi^0}(x)=
&+& \frac{1}{2\sqrt{6}f_\pi}\phi_S(x)\KET{6d_{\uparrow}u_{\downarrow}d_{\uparrow}+d_{\uparrow}d_{\downarrow}u_{\uparrow}+u_{\uparrow}d_{\downarrow}d_{\uparrow}}
\nonumber \\
&-& \frac{1}{2\sqrt{2}f_\pi}\phi_A(x)\KET{d_{\uparrow}d_{\downarrow}u_{\uparrow}-u_{\uparrow}d_{\downarrow}d_{\uparrow}}
\end{eqnarray}
\begin{eqnarray}
\phi_{n\pi^+}(x)=
&+& \frac{1}{\sqrt{12}f_\pi}\phi_S(x)\KET{2u_{\uparrow}d_{\downarrow}u_{\uparrow}-3u_{\uparrow}u_{\downarrow}d_{\uparrow}-3d_{\uparrow}u_{\downarrow}u_{\uparrow}} \nonumber \\
&-& \frac{1}{2f_\pi}\phi_A(x)\KET{u_{\uparrow}u_{\downarrow}d_{\uparrow}-d_{\uparrow}u_{\downarrow}u_{\uparrow}}
\end{eqnarray}
\begin{eqnarray}
\phi_{p\pi^-}(x)=
&-& \frac{1}{\sqrt{12}f_\pi}\phi_S(x)\KET{2d_{\uparrow}u_{\downarrow}d_{\uparrow}-3d_{\uparrow}d_{\downarrow}u_{\uparrow}-3u_{\uparrow}d_{\downarrow}d_{\uparrow}} \nonumber \\
&+& \frac{1}{2f_\pi}\phi_A(x)\KET{d_{\uparrow}d_{\downarrow}u_{\uparrow}-u_{\uparrow}d_{\downarrow}d_{\uparrow}}.
\end{eqnarray}

Let us now combine the hard and soft parts in order to obtain expressions for the usual nucleon form factors. For convenience, we will omit the dependence on the quark momentum fractions if possible. Moreover, we introduce the notations: $\phi_{SS}=\phi_S\phi_S$, $\phi_{AA}=\phi_A\phi_A$ and $\phi_{AS}=(\phi_A\phi_S+\phi_S\phi_A)$. We will expand the proton magnetic form factor, the neutron magnetic form factor, the isovector axial-vector form factor and the isoscalar axial-vector form factor
\pagebreak[0]
\begin{eqnarray}
G_M^p(Q^2) &=& \int[\intd\!x][\intd\!y]\phi^{\ast}_p(y)T_H^{em}(x,y,Q^2)\phi_p(x)\nonumber \\
&=& \frac{16}{9}\frac{(4\pi\bar{\alpha}_s)^2}{Q^4}\int[\intd\!x][\intd\!y]
\bigg\{2T_1\phi_{SS}-\frac{2}{\sqrt{3}}T_1\phi_{AS}+\frac{2}{3}[T_1+2T_2]\phi_{AA}\bigg\}
\end{eqnarray}
\begin{eqnarray}
G_M^n(Q^2) &=& \int[\intd\!x][\intd\!y]\phi^{\ast}_n(y)T_H^{em}(x,y,Q^2)\phi_n(x)\nonumber \\
&=& \frac{16}{9}\frac{(4\pi\bar{\alpha}_s)^2}{Q^4}\int[\intd\!x][\intd\!y]
\bigg\{\frac{2}{3}[T_2-T_1]\phi_{SS}+\frac{2}{\sqrt{3}}T_1\phi_{AS}+\frac{2}{3}[T_1-T_2]\phi_{AA}\bigg\}
\end{eqnarray}
\begin{eqnarray}
G_A^v(Q^2) &=& \int[\intd\!x][\intd\!y]\phi^{\ast}_n(y)T_{H5}^{(V-)}(x,y,Q^2)\phi_p(x)\nonumber \\
&=& \int[\intd\!x][\intd\!y]\phi^{\ast}_p(y)T_{H5}^{(V+)}(x,y,Q^2)\phi_n(x)\nonumber \\
&=& \frac{16}{9}\frac{(4\pi\bar{\alpha}_s)^2}{Q^4}\int[\intd\!x][\intd\!y]
\bigg\{\frac{2}{3}[4T_1+T_2]\phi_{SS}-\frac{4}{\sqrt{3}}T_1\phi_{AS}-2T_2\phi_{AA}\bigg\}
\end{eqnarray}
\begin{eqnarray}
G_A^s(Q^2) &=& \int[\intd\!x][\intd\!y]\phi^{\ast}_p(y)T_{H5}^{(S)}(x,y,Q^2)\phi_p(x)\nonumber \\
&=& \int[\intd\!x][\intd\!y]\phi^{\ast}_n(y)T_{H5}^{(S)}(x,y,Q^2)\phi_n(x)\nonumber \\
&=& \frac{16}{9}\frac{(4\pi\bar{\alpha}_s)^2}{Q^4}\int[\intd\!x][\intd\!y]
\bigg\{2[2T_1-T_2]\phi_{SS}+2[2T_1-T_2]\phi_{AA}\bigg\}.
\end{eqnarray}

Finally, we will combine the hard and soft parts in order to obtain expressions for the nucleon to pion nucleon transition form factors. We will receive expansions for all transition form factors similar to nucleon form factors
\pagebreak[0]
\begin{equation}
\begin{split}
& \mathcal{A}(\gamma^\star p \rightarrow p\pi^0)(Q^2)=\int[\intd\!x][\intd\!y]\phi^{\ast}_{p\pi^0}(y)T_H^{em}(x,y,Q^2)\phi_p(x)= \\
& \frac{16}{9}\frac{(4\pi\bar{\alpha}_s)^2}{Q^4}\frac{1}{f_\pi}\int[\intd\!x][\intd\!y]
\bigg\{\frac{1}{9}[23T_1-8T_2]\phi_{SS}+\frac{1}{\sqrt{3}}T_1\phi_{AS}-\frac{1}{3}[T_1+2T_2]\phi_{AA}\bigg\}
\end{split}
\end{equation}

\begin{equation}
\begin{split}
& \mathcal{A}(\gamma^\star p \rightarrow n\pi^+)(Q^2)=\int[\intd\!x][\intd\!y]\phi^{\ast}_{n\pi^+}(y)T_H^{em}(x,y,Q^2)\phi_p(x)= \\
& \frac{16}{9}\frac{(4\pi\bar{\alpha}_s)^2}{Q^4}\frac{1}{\sqrt{2}f_\pi}\int[\intd\!x][\intd\!y]
\bigg\{\frac{2}{9}[11T_1+4T_2]\phi_{SS}-\frac{2}{\sqrt{3}}T_1\phi_{AS}-\frac{2}{3}[T_1+2T_2]\phi_{AA}\bigg\}
\end{split}
\end{equation}

\begin{equation}
\begin{split}
& \mathcal{A}(\gamma^\star n \rightarrow n\pi^0)(Q^2)=\int[\intd\!x][\intd\!y]\phi^{\ast}_{n\pi^0}(y)T_H^{em}(x,y,Q^2)\phi_n(x)= \\
& \frac{16}{9}\frac{(4\pi\bar{\alpha}_s)^2}{Q^4}\frac{1}{f_\pi}\int[\intd\!x][\intd\!y]
\bigg\{\frac{13}{9}[T_1-T_2]\phi_{SS}+\frac{1}{\sqrt{3}}T_1\phi_{AS}+\frac{1}{3}[T_1-T_2]\phi_{AA}\bigg\}
\end{split}
\end{equation}

\begin{equation}
\begin{split}
& \mathcal{A}(\gamma^\star n \rightarrow p\pi^-)(Q^2)=\int[\intd\!x][\intd\!y]\phi^{\ast}_{p\pi^-}(y)T_H^{em}(x,y,Q^2)\phi_n(x)= \\
& \frac{16}{9}\frac{(4\pi\bar{\alpha}_s)^2}{Q^4}\frac{1}{\sqrt{2}f_\pi}\int[\intd\!x][\intd\!y]
\bigg\{\frac{2}{9}[T_2-T_1]\phi_{SS}+\frac{2}{\sqrt{3}}T_1\phi_{AS}+\frac{2}{3}[T_2-T_1]\phi_{AA}\bigg\}.
\end{split}
\end{equation}

All considered types of form factors depend on combinations of $\phi_S$ and $\phi_A$. In order to express the transition form factors as functions of nucleon form factors, we have to constrain these distribution functions.

We already know from large $Q^2$ processes that $\phi_S$ dominates $\phi_A$. Therefore, we consider now the basic case that $\phi_S$ is arbitrary and $\phi_A$ can be neglected. In this case, one can express all transition form factors as functions of two nucleon form factors. It makes sense to apply the dominant nucleon form factors $G_M^p$ and $G_M^n$ for this expansion. These results are already obtained in \cite{Pobylitsa:2001}. One gets
\pagebreak[0]
\begin{eqnarray}
\mathcal{A}(\gamma^\star p \rightarrow p\pi^0) &=& \frac{1}{f_\pi}\bigg(\frac{5}{6}G_M^p-\frac{4}{3}G_M^n\bigg) \\
\mathcal{A}(\gamma^\star p \rightarrow n\pi^+) &=& \frac{1}{\sqrt{2}f_\pi}\bigg(\frac{5}{3}G_M^p+\frac{4}{3}G_M^n\bigg) \\
\mathcal{A}(\gamma^\star n \rightarrow n\pi^0) &=& \frac{1}{f_\pi}\bigg(-\frac{13}{6}G_M^n\bigg) \\
\mathcal{A}(\gamma^\star n \rightarrow p\pi^-) &=& \frac{1}{\sqrt{2}f_\pi}\bigg(\frac{1}{3}G_M^n\bigg).
\end{eqnarray}

Moving away from the large $Q^2$ region, one cannot longer neglect $\phi_A$. Both components must be considered as arbitrary, but we can assume that $\phi_A$ is still small. That means, we can neglect the $\phi_{AA}$ parts for the form factor reduction. In order to express all transition form factors as functions of nucleon form factors, one has to use three nucleon form factors. We will add the form factor $G_A^v$ to the previous form factors $G_M^p$ and $G_M^n$. We avoid to use the form factor $G_A^s$, because it is suppressed in experiments. Taking into account reasonable antisymmetric contributions, we predict a more realistic description of the process and convincing agreement with experimental data consequentially
\pagebreak[0]
\begin{eqnarray}
\mathcal{A}(\gamma^\star p \rightarrow p\pi^0) &=& \frac{1}{f_\pi}\bigg(\frac{25}{6}G_M^p+\frac{2}{3}G_M^n-2G_A^v\bigg) \\
\mathcal{A}(\gamma^\star p \rightarrow n\pi^+) &=& \frac{1}{\sqrt{2}f_\pi}\bigg(\frac{5}{6}G_M^p+\frac{5}{6}G_M^n+\frac{1}{2}G_A^v\bigg) \\
\mathcal{A}(\gamma^\star n \rightarrow n\pi^0) &=& \frac{1}{f_\pi}\bigg(\frac{10}{3}G_M^p-\frac{1}{6}G_M^n-2G_A^v\bigg) \\
\mathcal{A}(\gamma^\star n \rightarrow p\pi^-) &=& \frac{1}{\sqrt{2}f_\pi}\bigg(\frac{5}{6}G_M^p+\frac{5}{6}G_M^n-\frac{1}{2}G_A^v\bigg).
\end{eqnarray}

\section{Comparison and Conclusion}

Our final task is to compare the obtained results for the nucleon structure functions with experimental data. We have the ability to apply experimental values for usual nucleon form factors in order to predict values for nucleon to pion nucleon transition form factors and nucleon structure functions. Data for the proton magnetic form factor are given in \cite{Arnold:1986} and \cite{Sill:1993} and for the neutron magnetic form factor in \cite{Rock:1982} and \cite{Rock:1992}. Moreover, data for the isovector axial-vector form factor are given in \cite{Kitagaki:1983}. One also needs values for multiple constants. Therefore, we can recommend \cite{Nakamura:2010}.

When $\phi_A$ can be neglected, we apply the following experimental values for the required nucleon form factors: $Q^4G_M^p(Q^2)=(1.0\pm 0.1)\,\textnormal{GeV}^4$ and $Q^4G_M^n(Q^2)=-(0.5\pm 0.1)\,\textnormal{GeV}^4$. In the case that $\phi_A$ is small, we use: $Q^4G_M^p(Q^2)=(1.1\pm 0.1)\,\textnormal{GeV}^4$ and $Q^4G_M^n(Q^2)=-(0.5\pm 0.1)\,\textnormal{GeV}^4$ and also $Q^4G_A^v(Q^2)=(1.5\pm 0.1)\,\textnormal{GeV}^4$. The composition of the results will always be done for the case when $\phi_A$ can be neglected at first and for the case when $\phi_A$ is small at last.

Let us first apply the discussed connection between nucleon to pion nucleon transition form factors and usual nucleon form factors in order to obtain results for the transition form factors. These expressions depend on the pion decay constant given by $f_\pi\approx 92.4\,\textnormal{MeV}$
\pagebreak[0]
\begin{eqnarray}
Q^4\mathcal{A}(\gamma^\star p \rightarrow p\pi^0)(Q^2) &=& (16.2\pm 0.1)\enspace\textnormal{GeV}^3 \\
Q^4\mathcal{A}(\gamma^\star p \rightarrow n\pi^+)(Q^2) &=& (7.7\pm 0.1)\enspace\textnormal{GeV}^3 \\
Q^4\mathcal{A}(\gamma^\star n \rightarrow n\pi^0)(Q^2) &=& (11.7\pm 0.1)\enspace\textnormal{GeV}^3 \\
Q^4\mathcal{A}(\gamma^\star n \rightarrow p\pi^-)(Q^2) &=& (-1.3\pm 0.1)\enspace\textnormal{GeV}^3
\end{eqnarray}
\begin{eqnarray}
Q^4\mathcal{A}(\gamma^\star p \rightarrow p\pi^0)(Q^2) &=& (13.5\pm 0.1)\enspace\textnormal{GeV}^3 \\
Q^4\mathcal{A}(\gamma^\star p \rightarrow n\pi^+)(Q^2) &=& (9.6\pm 0.1)\enspace\textnormal{GeV}^3 \\
Q^4\mathcal{A}(\gamma^\star n \rightarrow n\pi^0)(Q^2) &=& (8.1\pm 0.1)\enspace\textnormal{GeV}^3 \\
Q^4\mathcal{A}(\gamma^\star n \rightarrow p\pi^-)(Q^2) &=& (-1.9\pm 0.1)\enspace\textnormal{GeV}^3.
\end{eqnarray}

These values are just interesting from theoretical point of view. We remind that the corresponding transition states are undetected in experiments.

At next, we consider the ratio between a neutron and a proton structure function of the same kind at the production threshold of the pion. At this limit, the ratio is identical for all types of nucleon structure functions $F$ and generated by the transition form factors
\pagebreak[0]
\begin{equation}
\lim_{W\rightarrow W_{th}}\frac{F^n(W,Q^2)}{F^p(W,Q^2)}=
\frac{\sum_{X=X^n}|\mathcal{A}(\gamma^\star n \rightarrow X)|^2}{\sum_{X=X^p}|\mathcal{A}(\gamma^\star p \rightarrow X)|^2}=0.4\pm 0.1
\end{equation}
\begin{equation}
\lim_{W\rightarrow W_{th}}\frac{F^n(W,Q^2)}{F^p(W,Q^2)}=
\frac{\sum_{X=X^n}|\mathcal{A}(\gamma^\star n \rightarrow X)|^2}{\sum_{X=X^p}|\mathcal{A}(\gamma^\star p \rightarrow X)|^2}=0.3\pm 0.1.
\end{equation}

The obtained ratios are in good agreement with the results obtained in \cite{Braun:2008} and the perturbative QCD scaling limit expectation of $3/7$ for
$x_B \rightarrow 1$, see \cite{Farrar:1975}.

Finally, we will integrate the nucleon structure functions in an area near the threshold in order to obtain results which can be compared with experimental data. The expressions depend on the axial coupling constant given by $g_A\approx 1.27$ and we have to apply values for the included masses
$m_N\approx 0.94\,\textnormal{GeV}$ and $m_\pi\approx 0.14\,\textnormal{GeV}$.

We received experimental data for the integrated dominant proton structure function $F_2^p(W,Q^2)$ in the near threshold region, see \cite{Bosted:1994}. They integrated the quantity $Q^6F_2^p(W,Q^2)$ in different areas near the threshold and as expected, the best approximation was obtained for the lowest area of $W^2$, where the following integration was evaluated
\begin{equation}
\int_{th}^{1.4}\intd\!W^2Q^6F_2^p(W,Q^2)=(0.10\pm 0.02)\enspace\textnormal{GeV}^8.
\end{equation}
Concerning the integrated dominant neutron structure function $F_2^n(W,Q^2)$, we have no experimental data for comparisons at the moment. For several reasons, the required measurement can be done much easier on the proton than on the neutron. Nevertheless, we decided to integrate all nucleon structure functions in the same area in order to make predictions for further experiments.
\pagebreak[0]
\begin{eqnarray}
\int_{th}^{1.4}\intd\!W^2Q^6F_1^p(W,Q^2) &=& +(0.06\pm 0.02)\enspace\textnormal{GeV}^8 \\
\int_{th}^{1.4}\intd\!W^2Q^6F_2^p(W,Q^2) &=& +(0.11\pm 0.02)\enspace\textnormal{GeV}^8 \\
\int_{th}^{1.4}\intd\!W^2Q^6G_1^p(W,Q^2) &=& +(0.06\pm 0.02)\enspace\textnormal{GeV}^8 \\
\int_{th}^{1.4}\intd\!W^2Q^6G_2^p(W,Q^2) &=& -(0.05\pm 0.02)\enspace\textnormal{GeV}^8
\end{eqnarray}
\begin{eqnarray}
\int_{th}^{1.4}\intd\!W^2Q^6F_1^n(W,Q^2) &=& +(0.02\pm 0.02)\enspace\textnormal{GeV}^8 \\
\int_{th}^{1.4}\intd\!W^2Q^6F_2^n(W,Q^2) &=& +(0.05\pm 0.02)\enspace\textnormal{GeV}^8 \\
\int_{th}^{1.4}\intd\!W^2Q^6G_1^n(W,Q^2) &=& +(0.02\pm 0.02)\enspace\textnormal{GeV}^8 \\
\int_{th}^{1.4}\intd\!W^2Q^6G_2^n(W,Q^2) &=& -(0.02\pm 0.02)\enspace\textnormal{GeV}^8.
\end{eqnarray}
\pagebreak[0]
We obtained a result for the integrated proton structure function $F_2^p(W,Q^2)$ which is certainly in good agreement with the experimental value. Moreover, our results for this quantity and the integrated neutron structure function $F_2^n(W,Q^2)$ are the same as in \cite{Pobylitsa:2001}.
\pagebreak[0]
\begin{eqnarray}
\int_{th}^{1.4}\intd\!W^2Q^6F_1^p(W,Q^2) &=& +(0.05\pm 0.02)\enspace\textnormal{GeV}^8 \\
\int_{th}^{1.4}\intd\!W^2Q^6F_2^p(W,Q^2) &=& +(0.10\pm 0.02)\enspace\textnormal{GeV}^8 \\
\int_{th}^{1.4}\intd\!W^2Q^6G_1^p(W,Q^2) &=& +(0.05\pm 0.02)\enspace\textnormal{GeV}^8 \\
\int_{th}^{1.4}\intd\!W^2Q^6G_2^p(W,Q^2) &=& -(0.04\pm 0.02)\enspace\textnormal{GeV}^8
\end{eqnarray}
\begin{eqnarray}
\int_{th}^{1.4}\intd\!W^2Q^6F_1^n(W,Q^2) &=& +(0.01\pm 0.02)\enspace\textnormal{GeV}^8 \\
\int_{th}^{1.4}\intd\!W^2Q^6F_2^n(W,Q^2) &=& +(0.03\pm 0.02)\enspace\textnormal{GeV}^8 \\
\int_{th}^{1.4}\intd\!W^2Q^6G_1^n(W,Q^2) &=& +(0.01\pm 0.02)\enspace\textnormal{GeV}^8 \\
\int_{th}^{1.4}\intd\!W^2Q^6G_2^n(W,Q^2) &=& -(0.01\pm 0.02)\enspace\textnormal{GeV}^8.
\end{eqnarray}
\pagebreak[0]
We obtained a result for the integrated proton structure function $F_2^p(W,Q^2)$ which is in complete agreement with the experimental value. We predicted a more realistic description of the discussed process by taking into account a reasonable antisymmetric contribution and the perfect agreement with the experimental data will certainly confirm our statement.

$\phantom{X}$

The author wants to thank Prof. Dr. M. V. Polyakov for initiating this work and for useful comments.

The work has been supported by BMBF grant 06BO9012.

\bibliographystyle{h-physrev3}
\bibliography{Literature}

\end{document}